\documentstyle[11pt,newpasp,twoside,epsf]{article}
\def\edcomment#1{\iffalse\marginpar{\raggedright\sl#1\/}\else\relax\fi}
\reversemarginpar

\begin{document}
\title{Properties of CO Molecular Gas in IR Luminous Galaxies}
 \author{Lihong Yao}
\affil{Department of Astronomy \& Astrophysics, University of Toronto,
Toronto, ON M5S 3H8, Canada}

\begin{abstract}
We present the properties of the $^{12}$CO(1-0) and (3-2) line
emission from the nuclei of 60 IR luminous SLUGS galaxies.
This subsample is flux limited at $S_{60 \mu m}$ $\ge$ 5.24 Jy 
with FIR luminosities mostly at $L_{FIR}$ $>$ 10$^{10}$ L$_{\odot}$. 
The emission line strengths of $^{12}$CO(1-0) and (3-2) transitions
were compared at a common resolution of $\sim$ 15$^{\prime\prime}$,
and the line ratios $r_{31}$ vary from 0.22 to 1.72 with a mean value 
of 0.66 for the sources observed, indicating a large spread of the 
degree of excitation of CO in the sample. Our analysis shows that 
(1) there is a non-linear relation between CO and FIR luminosities, such 
that their ratio $L_{CO}$/$L_{FIR}$ decreases linearly with increasing 
$L_{FIR}$, (2) we find a possible dependence of the degree of 
CO gas excitation on the efficiency of star forming activity, 
(3) using the large velocity gradient (LVG) approximation to model 
the observed data, the results show that the mean value of the CO-to-H$_2$ 
conversion factor $X$ for the SLUGS sample is lower by a factor of 10 
compared to the conventional value derived for the Galaxy, assuming 
that the abundance of CO relative to H$_2$ is 10$^{-4}$, 
(4) due to a contribution to the SCUBA brightness measurement by 
$^{12}$CO(3-2) emission, the average dust mass is reduced by 25-38\% , 
and the mean global gas-to-dust mass ratio is reduced from 430 
to 360, but is further reduced to 100 when applied to 
the nuclear regions of the SLUGS galaxies, (5) for a subset of 
12 galaxies with H I maps, we derive a mean total face-on surface 
density of H$_2$+H I of about 42 M$_{\odot}$ pc$^{-2}$ within $\sim$ 
2 kpc of the nucleus, which is intermediate between that in 
galaxies like our own and those with strong star formation.
\end{abstract}

\section{Introduction} 

Understanding the properties and evolution of the gas and dust content in 
nearby IR luminous galaxies is essential for understanding the cause
and temporal evolution of starburst activity and its role in the
cosmic evolution of galaxies. Therefore, there is a need to investigate 
large statistical samples of IR luminous galaxies using a multitude 
of different types of data, including CO, H I, and continuum in the 
submillimeter (sub-mm), FIR and radio, in order to constrain theories 
of how the interstellar medium (ISM) evolves. Millimeter and submillimeter 
CO rotational lines are often used as tracers of molecular hydrogen. 
The ratio of $^{12}$CO(3-2) to (1-0) line emission provides more sensitive 
measure of gas temperature and density than the ratio of $^{12}$CO(2-1) to 
(1-0) lines. Recently, we have presented the largest statistical CO 
survey (Yao et al. 2003) for the nearby universe by investigating the 
properties of CO molecular gas in a nearly complete subsample of 60 
IR luminous galaxies selected from the SLUGS Survey (Dunne et al. 2000).
The SLUGS survey containing 104 galaxies is based on the Revised Bright 
Galaxy Sample of {\it IRAS} galaxies (Soifer et al. 1987) within 
-10$^{\circ}$ $\le$ $\delta$ $\le$ +50$^{\circ}$ and with $cz$ $\ge$ 
1900 km s$^{-1}$, a flux limit of $S_{60\mu m}$ $\ge$ 5.24 Jy, and 
the FIR luminosity $L_{FIR}$ $\ge$ 10$^{10}$ L$_{\odot}$. 

\section{Properties of Molecular Gas and Dust}

A study of molecular gas of a complete flux-limited subsample is important 
to complement the study of the dust. The angular resolution of the
$^{12}$CO(1-0) and (3-2) point observations presented in our CO survey are 
nearly identical ($\sim$ 15$^\prime$$^\prime$). A complete discussion 
of the observations, the CO line spectra, and the data analysis is contained 
in Yao et al. (2003) and references therein. Together with the sub-mm 
fluxes (Dunne et al. 2000), plus existing data on H I (Thomas et al. 2002), 
radio continuum (Condon et al. 1990), and FIR (Dunne et al. 2000; 
NED\footnote{The NASA/IPAC Extragalactic Database operated by the Jet 
propulsion Laboratory.}), we are able to search for a relationship 
between the degree of excitation of the CO in this sample and the star 
forming properties.

\begin{figure}
\plotone{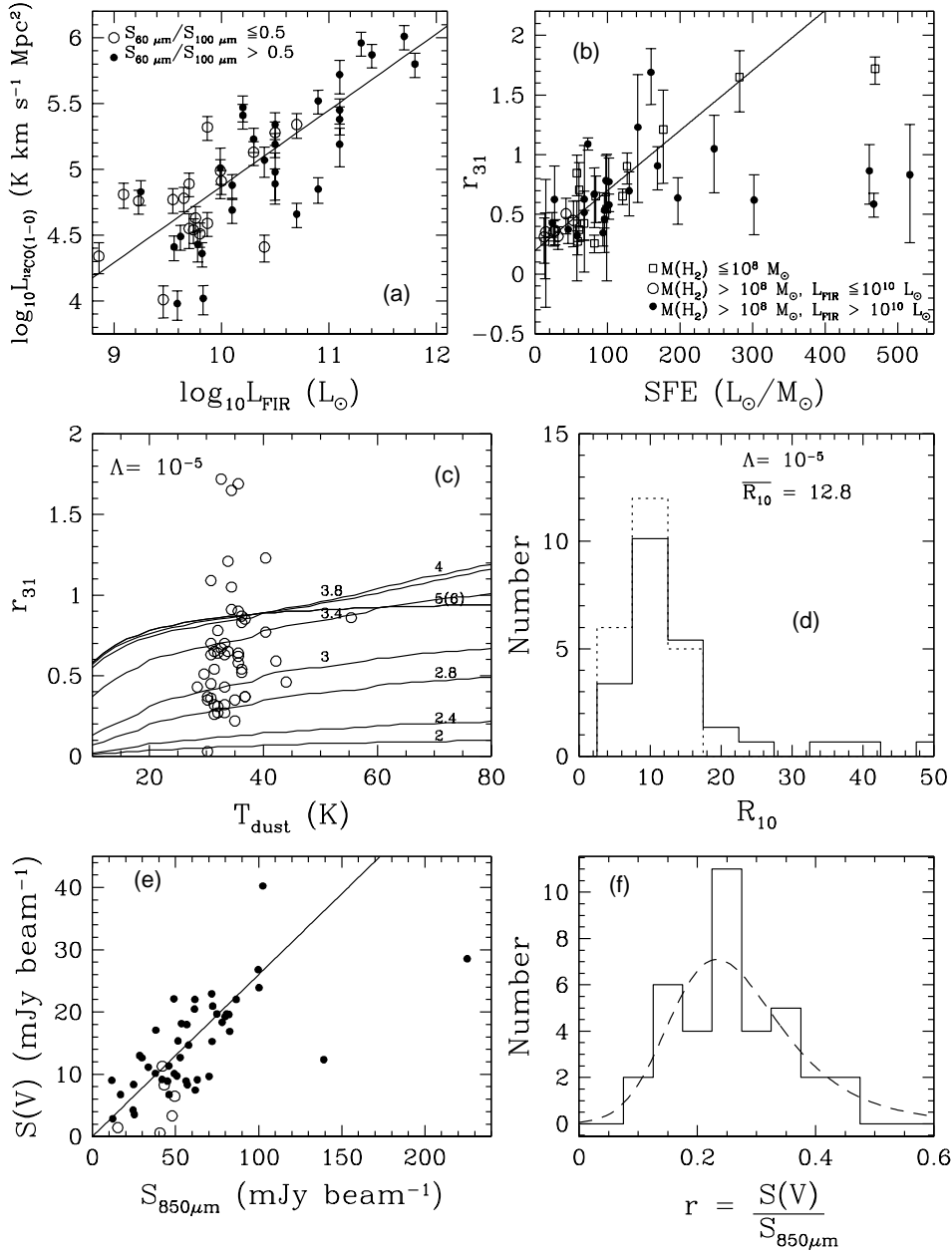}
\caption{(a) CO luminosity vs. FIR luminosity applicable to the 
15$^{\prime\prime}$ beam, (b) the $r_{31}$ ratio vs. star formation 
efficiency, the line represents a linear regression fit to the data
with SFE $\le$ 200 L$_{\odot}$/M$_{\odot}$, the $M$(H$_2$) is computed
using our derived $X$-factor, i.e. 2.7$\times$10$^{19}$ 
cm$^{-2}$(K km s$^{-1}$)$^{-1}$, (c) $r_{31}$ vs. $T_{dust}$ for
different H$_2$ gas densities log$_{10}$ $n$(H$_2$) based on LVG model 
for $\Lambda$ = 10$^{-5}$ (km s$^{-1}$ pc$^{-1}$), assuming 
$T_{kin}$ = $T_{dust}$, the $^{12}$CO abundance is 10$^{-4}$, 
and the isotope abundance ratio $^{12}$CO/$^{13}$CO = 40, (d) a
comparison between the distributions of computed (dotted line) 
and observed isotope $^{12}$CO(1-0)/$^{13}$CO(1-0) ratios (solid line) 
for the best agreement regime of $\Lambda$ = 10$^{-5}$, (e) SCUBA equivalent 
flux $S$($V$) produced by $^{12}$CO(3-2) line vs. SCUBA flux at the central 
pixel for SLUGS sources for $v$ $<$ 10,000 km s$^{-1}$ (filled circles) and 
$v$ $>$ 10,0000 km s$^{-1}$ (open circles), the slope for the linear regression 
fit is 0.26$\pm$0.01, and (f) histogram of $S$($V$), the dash curve represents
the expected distribution for a true ratio of $r$ = 0.24 and mean fractional
uncertainties of respectively 32\% and 25\% for $S$($V$) (numerator) 
and $S_{850\mu m}$ (denominator).}
\end{figure}

\subsection{CO and FIR Luminosities}

Figure 1(a) shows a comparison between the CO and FIR luminosities
for two different FIR color regimes. A linear fit to the 1-0 data in the 
log-log plane yields log$_{10} L_{^{12}CO(1-0)}$ = (0.2 $\pm$ 0.7) + 
(0.57 $\pm$ 0.07) log$_{10} L_{FIR}$. The nonlinear variation of $L_{CO}$ 
with $L_{FIR}$ implies a decrease in the ratio $L_{CO}$/$L_{FIR}$ with 
increasing $L_{FIR}$, so that more IR luminous galaxies have higher 
dust temperatures and star formation efficiency. No correlation with 
the projected beam size on the galaxies is evident. The ratio depends 
intrinsically on the total FIR luminosity. This implies that for high-$z$
objects where $L_{FIR}$ is exceedingly high, the corresponding $L_{CO}$ 
will be seriously overestimated if a linear relation between $L_{CO}$
and $L_{FIR}$ is assumed.  

\subsection{Star Formation Efficiency}

We examined the excitation of CO and its relation with the properties 
of gas/dust and star formation in the the central starburst regions
in the SLUGS sample. There are no significant correlations between
the $^{12}$CO(3-2) to (1-0) line ratio $r_{31}$ and the distance of
galaxies, star formation rate, $T_{dust}$ and $M_{dust}$, H$_2$ gas mass,
and the color indices. This might reflect a range of localized conditions 
in the molecular clouds. But the degree of CO excitation measured by 
$r_{31}$ varies greatly from galaxy to galaxy. There is a trend for
the $r_{31}$ to increase with increasing concentration and efficiency 
of star-forming activity (see Figure 1 (b)). The saturation of $r_{31}$
seen at higher SFE implies that the gas is denser and warmer in 
regions of higher SFE, which is consistent with the nonlinearity 
between $L_{CO}$ and $L_{FIR}$.

\subsection{The CO-to-H$_2$ Conversion Factor $X$}

Our CO line measurements, together with dust and CO isotope data
taken from the literature (Dunne et al. 2000; Aalto et al. 1995; 
Taniguchi et al. 1999), are modeled using the LVG approximation 
yielding $X$ = $\frac{n(H_2) \Lambda}{X_{CO} T_{rad}}$ 
(see Figure 1 (c) and (d)) to estimate that the controversial 
$X$ lies between 1.3$\times$10$^{19}$ and 6.7$\times$10$^{19}$ 
cm$^{-2}$(K km s$^{-1}$)$^{-1}$ for SLUGS galaxies, which is 
about 4-20 times lower than the conventional $X$ factor derived 
from the Galaxy. The mean value 2.7$\times$10$^{19}$ is comparable 
to that estimated from diffuse clouds in the Galaxy and with that 
found for extreme starbursts in nearby galaxies. 
 
\subsection{Virial Stability of the Molecular Clouds}

Using virial analysis, we show that the molecular clouds in starburst
regions are not gravitationally bound, confirming the suggestion 
recently made by Zhu et al. (2003) for the Antennae galaxies, 
unless one is willing to accept a 9-90 times lower [CO/H$_2$] 
abundance ratio. The cause could be the destruction by the stellar 
winds and expanding shells of newborn massive stars. Most of the 
CO line emission originates from the nonvirialized warm and 
diffuse gas clouds.

\subsection{Revised Masses of Dust and Gas of SLUGS Galaxies}

Our $^{12}$CO(3-2) observations provide an important database for 
correcting the SCUBA 850 $\mu$m data in the SLUGS sample (see Figure 1 (e)
and (f)), thus permitting a revised characterization of the masses of 
dust and gas (see Seaquist et al. 2003). The average downward correction 
of dust mass is 25-38\%, which has no bearing on earlier conclusions 
concerning the shapes of the dust mass-luminosity function derived 
from SLUGS survey. Using the $^{12}$CO(3-2)/(1-0) ratios in Yao et al. 
(2003), we estimate that it is unlikely that any correction is required
for the contribution by $^{12}$CO(6-5) to 450 $\mu$m fluxes of the 
SLUGS galaxies measured with SCUBA. The revised mean gas-to-dust ratio
(g/d) is reduced from 310 to 100, which is significantly lower than 
those reported for the global g/d ratios for IR luminous galaxies. 
 
\acknowledgements I thank my advisor Prof. Ernie Seaquist and
Reinhardt Travel Fellowship at Univ. of Toronto for their support. I also 
thank Kavli Institute for Theoretical Physics for their generosity.

\end{document}